# The effect of tensor interaction in splitting the energy levels of relativistic systems

Mohammad Reza Shojaei[1] and Mohsen Mousavi[2]

[1, 2] Department of physics, University of Shahrood, P.O. Box 36155-316, Shahrood, Iran

[1]mahyashojaei@yahoo.com

[2]nuclear.physics2020@gmail.com

**Abstract:** In this paper we solve approximately Dirac equation for Eckart plus Hulthen potentials with Coulomb-like and Yukawa-like tensor interaction in the presence of Spin and Pseudo-spin Symmetry for k≠0. The formula method is use to obtain the energy Eigen-values and wave functions. We also discuss the energy Eigen-values and the wave function of the radial Dirac equation for Eckart plus Hulthen potentials with formula method. To show the accuracy of the present model, some numerical results are shown in both pseudo-spin and spin symmetry limits.

## 1. Introduction

One of the interesting problems in nuclear and high energy physics is to obtain analytical solution of the Klein - Gordon, Duffin–Kemmer - Petiau and Dirac equations for mixed vector and scalar potentials [1]. The study of relativistic effect is always useful in some quantum mechanical systems [2, 3]. Therefore, the Dirac equation has become the most appealing relativistic wave equation for spin-1/2 particles. For example, in the relativistic treatment of nuclear phenomena, the Dirac equation is use to describe the behavior of the nuclei in nucleus and also to solve many problems of high-energy physics and chemistry. For this reason, it has been used extensively to study the relativistic heavy ion collisions, heavy ion spectroscopy and more recently in laser–matter interaction (for a review, see [4] and references there) and condensed matter physics [5, 6]. The idea about spin symmetry and pseudo-spin symmetry with the nuclear shell model have been introduced in 1969 by Arima et al. (1969), Hecht and Adler (1969) [7, 8]. Spin and pseudo-spin symmetries are SU (2) symmetries of a Dirac Hamiltonian with vector and scalar potentials. They are realized when the difference, Δ(r) =V(r) −S(r), or the sum, Σ(r) =V(r) +S(r), are constant. The near realization of these symmetries may explain degeneracy in some heavy meson spectra (spin symmetry) or in single-particle energy levels in nuclei (pseudo-spin symmetry), when these physical systems are described by relativistic mean-field theories (RMF) with scalar and vector potentials [9]. Recently, some authors have studied various types of potential with a tensor potential, under the conditions of pseudo-spin and spin symmetry. They have found out that the tensor interaction removes the

١

degeneracy between two states in the pseudo-spin and spin doublet [10, 11, 12, 13]. The pseudo-spin and spin symmetry appeared in nuclear physics refers to a quasi-degeneracy of the single-nucleon doublets and can be characterized with the non-relativistic quantum numbers (n, l, j = l + 1/2) and (n, l + 2, j = l + 3/2), where n, l and j are the single-nucleon radial, orbital and total angular momentum quantum numbers for a single particle, respectively [7, 8]. The kind of various methods have been used for the analytical solutions of the Klein–Gordon equation and Dirac equation such as the Super symmetric Quantum Mechanics [14, 15, 16], Asymptotic Iteration method (AIM) [17, 18], factorization method [19, 20], Laplace transform approach [21], GPS method [22, 23] and the path integral method [24, 25, 26], Nikiforov-Uvarov method [27, 28, 29] and others. The Klein – Gordon and Dirac wave equations are frequently used to describe the particle dynamics in relativistic quantum mechanics with some typical kinds of potential by using different methods [30]. For example, Kratzer potential [31, 32], Woods-Saxon potential [33, 34], Scarf potential [35, 36], Hartmann potential [37, 38], Rosen Morse potential,[39, 40], Hulthen potential [41] and Eckart potential [42, 43].

In this paper, we attempt to solve approximately Dirac wave equation for k≠0 for Eckart plus Hulthen potentials for the spin and pseudo-spin symmetry with a tensor potential by using the Formula method. The organization of this paper is as follows: in Section 2, the Formula method is reviewed [44]. In section 3 we review Basic Dirac Equations briefly. In section 4.1 and 4.2, solutions of Dirac wave equation for the spin and pseudo-spin symmetry of these potentials in the presence of Coulomb-like tensor interaction are presented, respectively. In section 5.1 and 5.2 solutions of Dirac wave equation for the spin and pseudo-spin symmetry of these potentials in the presence of Coulomb-like plus Yukawa-like tensor interaction are presented, respectively. In section 6 we provide results and discussion. The conclusion is given in Section 7.

## 2. Review of Formula Method

The Formula method has been used to solve the Schrodinger, Dirac and Klein-Gordon wave equations for a certain kind of potential. In this method the differential equations can be written as follows [44]:

$$\Psi''_n(s) + \frac{(k_1 - k_2 s)}{s(1 - k_3 s)} \Psi'_n(s) + \frac{(As^2 + Bs + C)}{s^2 (1 - k_3 s)^2} \Psi_n(s) = 0 \qquad (1)$$

For a given Schrödinger-like equation in the presence of any potential model which can be written in the form of Eq. (1), the energy Eigen-values and the corresponding wave function can be obtained by using the following formulas respectively [44]:

٢

$$\left[\dfrac{k_4^2-k_5^2-\left[\dfrac{1-2n}{2}-\dfrac{1}{2k_3}\left(k_2-\sqrt{(k_3-k_2)^2-4A}\right)\right]^2}{2\left[\dfrac{1-2n}{2}-\dfrac{1}{2k_3}\left(k_2-\sqrt{(k_3-k_2)^2-4A}\right)\right]}\right]^2-k_5^2=0, k_3\neq 0 \qquad (2)$$

$$\Psi_n(s)=N_n s^{k_4}(1-k_3 s)^{k_5}{}_2F_1\left(-n, n+2(k_4+k_5)+\dfrac{k_2}{k_3}-1; 2k_4+k_1, k_3 s\right) \qquad (3)$$

Where,

$$k_4=\dfrac{(1-k_1)+\sqrt{(1-k_1)^2-4C}}{2}$$

$$k_5=\dfrac{1}{2}+\dfrac{k_1}{2}-\dfrac{k_2}{2k_3}+\sqrt{\left[\dfrac{1}{2}+\dfrac{k_1}{2}-\dfrac{k_2}{2k_3}\right]^2-\left[\dfrac{A}{k_3^2}+\dfrac{B}{k_3}+C\right]} \qquad (4)$$

And $N_n$ is the normalization constant. In special case where $k_3 \to 0$ the energy Eigen-values and the corresponding wave function can be obtained as [44]:

$$\left[\dfrac{B-k_4 k_2-n k_2}{2k_4+k_1+2n}\right]^2-k_5^2=0 \qquad (5)$$

$$\Psi_n(s)=N_n s^{k_4}\exp(-k_5 s){}_1F_1\left(-n; 2k_4+k_1; (2k_5+k_2)s\right) \qquad (6)$$

The solution provides a valuable means for checking and improving models and numerical methods introduced for solving complicated quantum systems.

## 3. Basic Dirac Equations

In the relativistic description, the Dirac equation of a single-nucleon with the mass moving in an attractive scalar potential S(r) and a repulsive vector potential V(r) can be written as [45]

$$[-i\hbar c\hat{\alpha}.\hat{\nabla}+\hat{\beta}(Mc^2+S(r))]\Psi_{n_r,k}=[E-V(r)]\Psi_{n_r,k} \qquad (7)$$

Where $E$ is the relativistic energy, $M$ is the mass of a single particle and α and β are the 4×4 Dirac matrices. For a particle in a central field, the total angular momentum J and $\hat{K}=-\hat{\beta}(\hat{\alpha}.\hat{L}+\hbar)$ commute with the Dirac Hamiltonian where L is the orbital angular momentum. For a given total angular momentum j, the Eigen-values of the $\hat{K}$ are k=±(j+1/2) where negative sign is for aligned spin and positive sign is for unaligned spin. The wave-functions can be classified according to their angular momentum *j* and spin-orbit quantum number k as follows:

$$\Psi_{n_r,k}(r,\theta,\phi) = \frac{1}{r}\begin{bmatrix} F_{n_r,k}(r)Y_{jm}^l(\theta,\phi) \\ iG_{n_r,k}(r)Y_{jm}^{\tilde{l}}(\theta,\phi) \end{bmatrix} \quad (8)$$

Where $F_{n_r,k}(r)$ and $G_{n_r,k}(r)$ are upper and lower components, $Y_{jm}^l(\theta,\phi)$ and $Y_{jm}^{\tilde{l}}(\theta,\phi)$ are the spherical harmonic functions. $n_r$ is the radial quantum number and m is the projection of the angular momentum on the z axis. The orbital angular momentum quantum numbers l and $\tilde{l}$ represent to the spin and pseudo-spin quantum numbers. Substituting Eq. (8) into Eq. (7), we obtain couple equations for the radial part of the Dirac equation as follows by $\hbar=c=1$

$$\begin{cases} \left(\dfrac{d}{dr}+\dfrac{k}{r}-U(r)\right)F_{n_r,k}(r) = [M+E_{n,k}-\Delta(r)]G_{n_r,k}(r) \\ \left(\dfrac{d}{dr}-\dfrac{k}{r}+U(r)\right)G_{n_r,k}(r) = [M-E_{n,k}+\Sigma(r)]F_{n_r,k}(r) \end{cases} \quad (9)$$

Where $\Delta(r)=V(r)-S(r)$ and $\Sigma(r)=V(r)+S(r)$ are the difference and the sum of the potentials V(r) and S(r), respectively and U(r) is a tensor potential. We obtain the second-order Schrodinger-like equation as:

$$\left\{\dfrac{d^2}{dr^2}-\dfrac{k(k+1)}{r^2}+\dfrac{2kU(r)}{r}-\dfrac{dU(r)}{dr}-U^2(r)-[M+E_{n,k}-\Delta(r)][M-E_{n,k}+\Sigma(r)] \right. \\ \left. +\dfrac{\dfrac{d\Delta(r)}{dr}\left(\dfrac{d}{dr}+\dfrac{k}{r}-U(r)\right)}{(M+E_{n,k}-\Delta(r))}\right\}F_{n_r,k}(r)=0 \quad (10)$$

$$\left\{\dfrac{d^2}{dr^2}-\dfrac{k(k-1)}{r^2}+\dfrac{2kU(r)}{r}+\dfrac{dU(r)}{dr}-U^2(r)-[M+E_{n,k}-\Delta(r)][M-E+\Sigma(r)] \right. \\ \left. +\dfrac{\dfrac{d\Sigma(r)}{dr}\left(\dfrac{d}{dr}-\dfrac{k}{r}+U(r)\right)}{(M-E_{n,k}+\Sigma(r))}\right\}G_{n_r,k}(r)=0 \quad (11)$$

We consider bound state solutions that demand the radial components satisfying $F_{n_r,k}(0)=G_{n_r,k}(0)=0$, and $F_{n_r,k}(\infty)=G_{n_r,k}(\infty)=0$ [45].

### 4.1. Solution Spin Symmetric with Coulomb-like tensor interaction

Under the condition of the spin symmetry, i. e. $\dfrac{d\Delta(r)}{dr}=0$ or $\Delta(r) = C_s$=const, the upper component Dirac equation can be written as:

$$\left\{\dfrac{d^2}{dr^2}-\dfrac{k(k+1)}{r^2}+\dfrac{2kU(r)}{r}-\dfrac{dU(r)}{dr}-U^2(r)-[M+E_{n,k}-C_s][M-E_{n,k}+\Sigma(r)]\right\}F_{n_r,k}(r)=0 \quad (12)$$

The potential $\Sigma(r)$ is taken as the Eckart [42, 43] plus Hulthen potentials [41]

۴

$$\Sigma(r) = 4q_1 \frac{e^{-2\alpha r}}{(1-e^{-2\alpha r})^2} - q_2 \frac{(1+e^{-2\alpha r})}{(1-e^{-2\alpha r})} + \frac{v_0}{(1-e^{-2\alpha r})} - \frac{v_1}{(1-e^{-2\alpha r})^2} \qquad (13)$$

Where the parameters $q_1$, $q_2$, $v_0$ and $v_1$ are real parameters, these parameters describe the depth of the potential well, and the parameter $\alpha$ is related to the range of the potential.

For the tensor term, we consider the Coulomb-like potential [46],

$$U(r) = -\frac{H}{r}, \quad H = \frac{Z_f Z_g e^2}{4\pi\varepsilon_0}, \quad r \rangle R_c \qquad (14)$$

Where $R_c$ is the coulomb radius, $Z_f$ and $Z_g$ stand for the charges of the projectile particle f and the target nucleus g, respectively.

By substituting Eq. (13) and (14) into Eq. (12), we obtain the upper radial equation of Dirac equation as:

$$\left\{ \frac{d^2}{dr^2} - \frac{\tau_k(\tau_k+1)}{r^2} - \gamma - \beta\left( 4q_1 \frac{e^{-2\alpha r}}{(1-e^{-2\alpha r})^2} - q_2 \frac{(1+e^{-2\alpha r})}{(1-e^{-2\alpha r})} + \frac{v_0}{(1-e^{-2\alpha r})} - \frac{v_1}{(1-e^{-2\alpha r})^2} \right) \right\} F_{n_r,k}(r) = 0 \qquad (15)$$

Where $\tau_k = k+H$, $\gamma = (M+E_{n,k}-C_s)(M-E_{n,k})$ and $\beta = (M+E_{n,k}-C_s)$.

Equation (15) is exactly solvable only for the case of k = 0. In order to obtain the analytical solutions of Eq. (15), we employ the improved approximation scheme suggested by Greene and Aldrich [47, 48] and replace the spin–orbit coupling term with the expression that is valid for $\alpha \leq 1$ [49].

$$\frac{1}{r^2} \approx \frac{4\alpha^2}{(1-e^{-2\alpha r})^2} \qquad (16)$$

The behavior of the improved approximation is plotted in Fig. 1. We can see the good agreement for small $\alpha$ values.

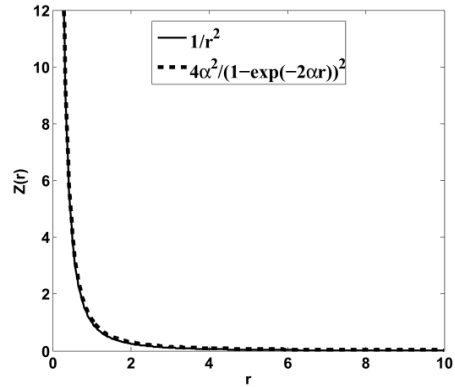

**Fig. 1.** The behavior approximation for $\alpha = 0.07$ fm$^{-1}$.

By applying the transformation $s = \exp(-2\alpha r)$ Eq. (15) brings into the form:

٥

$$F''_{n,k}(s) + \frac{(1-s)}{s(1-s)} F'_{n,k}(s) + \frac{1}{s^2(1-s)^2}\left[As^2 + Bs + C\right]F_{n,k}(s) = 0 \tag{17}$$

Where the parameters A, B and C are considered as follows:

$$\begin{aligned}A &= -\frac{1}{4\alpha^2}\left[\gamma + \beta q_2\right] \\ B &= \frac{1}{4\alpha^2}\left[2\gamma - 4\beta q_1 + \beta v_0\right] \\ C &= -\frac{1}{4\alpha^2}\left[4\alpha^2\tau_k(\tau_k - 1) + \gamma - \beta q_2 + \beta v_0 - \beta v_1\right]\end{aligned} \tag{18}$$

Now by comparing Eq. (17) with Eq. (1), we can easily obtain the coefficients $k_i$ (i = 1, 2, 3) as follows:

$$k_1 = k_2 = k_3 = 1 \tag{19}$$

The values of the coefficients $k_i$ (i = 4, 5) are also found from Eq. (4) as below:

$$\begin{aligned}k_4 &= \sqrt{-C} \\ k_5 &= \frac{1}{2} + \sqrt{\frac{1}{4} - [A+B+C]}\end{aligned} \tag{20}$$

Thus, by the use of energy equation Eq. (2) for energy Eigen-values, we find:

$$\left[\frac{-C - \left[\frac{1}{2} + \sqrt{\frac{1}{4} - [A+B+C]}\right]^2 - \left[\frac{1-2n}{2} - \frac{1}{2}\left(1 - \sqrt{-4A}\right)\right]^2}{2\left[\frac{1-2n}{2} - \frac{1}{2}\left(1 - \sqrt{-4A}\right)\right]}\right]^2 - \left[\frac{1}{2} + \sqrt{\frac{1}{4} - [A+B+C]}\right]^2 = 0 \tag{21}$$

In Tables 1–3, we give the numerical results for the spin symmetric energy Eigen-values (in units of fm$^{-1}$).

**Table 1.** Energies of the spin symmetry limit in the presence and absence of Coulomb-like tensor interaction by parameters M=10 fm$^{-1}$, c=1, h=1, α=0.05fm$^{-1}$, V$_1$=0.4fm$^{-1}$, V$_0$=0.3fm$^{-1}$, q$_1$=0.1fm$^{-1}$, q$_2$=0.2fm$^{-1}$, C$_s$=5fm$^{-1}$

| l | n,k>0 | State(l,j) | E$^s_{n,k}$(H=0) | E$^s_{n,k}$(H=0.65) | n,k<0 | State(l,j+1) | E$^s_{n,k}$(H=0) | E$^s_{n,k}$(H=0.65) |
|---|---|---|---|---|---|---|---|---|
| 1 | 1,1 | 1p$_{1/2}$ | 3.854830541 | 5.230932502 | 1,-2 | 1p$_{3/2}$ | 3.854830541 | 2.022333483 |
| 2 | 1,2 | 1d$_{3/2}$ | 5.816938791 | 6.683341758 | 1,-3 | 1d$_{5/2}$ | 5.816938791 | 4.646742373 |
| 3 | 1,3 | 1f$_{5/2}$ | 7.054878462 | 7.612226007 | 1,-4 | 1f$_{7/2}$ | 7.054878462 | 6.315225843 |
| 4 | 1,4 | 1g$_{7/2}$ | 7.855626422 | 8.227670602 | 1,-5 | 1g$_{9/2}$ | 7.855626422 | 7.373922639 |
| 1 | 2,1 | 2p$_{1/2}$ | 5.770989098 | 6.634606185 | 2,-2 | 2p$_{3/2}$ | 5.770989098 | 4.607350329 |
| 2 | 2,2 | 2d$_{3/2}$ | 7.005594455 | 7.562996503 | 2,-3 | 2d$_{5/2}$ | 7.005594455 | 6.267433123 |
| 3 | 2,3 | 2f$_{5/2}$ | 7.806806149 | 8.180021794 | 2,-4 | 2f$_{7/2}$ | 7.806806149 | 7.324529686 |
| 4 | 2,4 | 2g$_{7/2}$ | 8.346470004 | 8.605903663 | 2,-5 | 2g$_{9/2}$ | 8.346470004 | 8.019210395 |

**Table 2.** The energy Eigen-values (in units of fm$^{-1}$) for the spin symmetry limit with parameters M=10fm$^{-1}$, c=1, h=1, V$_1$=4fm$^{-1}$, V$_0$=3fm$^{-1}$, q$_1$=1fm$^{-1}$, q$_2$=2fm$^{-1}$, C$_s$ =5fm$^{-1}$.

| α(fm⁻¹) | $E_{n,k}(H=0)$ | | $E_{n,k}(H=0.65)$ | |
|---|---|---|---|---|
| | $1d_{3/2}$ | $1d_{5/2}$ | $1d_{3/2}$ | $1d_{5/2}$ |
| 0.3 | 2.367211785 | 2.367211785 | 3.624581074 | 0.93609384 |
| 0.35 | 3.53989514 | 3.53989514 | 4.771398349 | 2.070547382 |
| 0.4 | 4.521194951 | 4.521194951 | 5.690996586 | 3.069929104 |
| 0.5 | 5.998968975 | 5.998968975 | 7.009500454 | 4.671333928 |
| 0.6 | 6.995852039 | 6.995852039 | 7.853318047 | 5.827299953 |
| 0.7 | 7.670572507 | 7.670572507 | 8.400594789 | 6.652859598 |

**Table 3.** The energy Eigen-values (in units of fm−1) for the spin symmetry limit with parameters c=1, h=1, α=0.05fm⁻¹, V₁=4fm⁻¹, V₀=3fm⁻¹, q₁=1fm⁻¹, q₂=2fm⁻¹, C$_s$ =5fm⁻¹

| M(fm⁻¹) | $E_{n,k}(H=0)$ | | $E_{n,k}(H=0.65)$ | |
|---|---|---|---|---|
| | $1d_{3/2}$ | $1d_{5/2}$ | $1d_{3/2}$ | $1d_{5/2}$ |
| 5.5 | 3.335677895 | 3.335677895 | 3.893525562 | 2.662575645 |
| 6 | 3.479942785 | 3.479942785 | 4.105505756 | 2.72071499 |
| 6.5 | 3.618211096 | 3.618211096 | 4.311706034 | 2.7725715 |
| 7 | 3.752572889 | 3.752572889 | 4.514086551 | 2.820419493 |
| 8 | 4.013951033 | 4.013951033 | 4.911575769 | 2.908709541 |
| 9 | 4.269356779 | 4.269356779 | 5.303088298 | 2.991067928 |
| 10 | 4.521194951 | 4.521194951 | 5.690996586 | 3.069929104 |

Let us find the corresponding wave functions. In reference to Eq. (3) and Eq. (20), we can obtain the upper wave function as:

$$F_{n,k}^s(r) = N\left(e^{-2\alpha r}\right)^{\left(\sqrt{-c}\right)}\left(1-e^{-2\alpha r}\right)^{\left(\frac{1}{2}+\sqrt{\frac{1}{4}+A+B+C}\right)} {}_2F_1\left(-n, n+2\left(\sqrt{-c}+\frac{1}{2}+\sqrt{\frac{1}{4}+A+B+C}\right); 2\sqrt{-c}+1, e^{-2\alpha r}\right) \quad (22)$$

Where N is the normalization constant, on the other hand, the lower component of the Dirac spinor can be calculated from Eq. (23) as:

$$G_{n,k}^s(r) = \frac{1}{M+E_{n,k}^s - C_s}\left(\frac{d}{dr}+\frac{k}{r}-U(r)\right)F_{n,k}^s(r) \quad (23)$$

We have obtained the energy Eigen-values and the wave function of the radial Dirac equation for Eckart plus Hulthen potentials with Coulomb-like tensor interaction in the presence of the spin symmetry for k≠0.

### 4.2. Solution Pseudo-spin Symmetric with Coulomb-like tensor interaction

For pseudo-spin symmetry, i.e., $\frac{d\Sigma(r)}{dr}=0$ or $\Sigma(r) = C_{ps}$=const the lower component Dirac equation can be written as:

$$\left\{\frac{d^2}{dr^2}-\frac{k(k-1)}{r^2}+\frac{2kU(r)}{r}+\frac{dU(r)}{r}-U^2(r)-[M+E_{n,k}-\Delta(r)][M-E+\Sigma(r)]\right\}G_{n_r,k}(r)=0 \quad (24)$$

We consider the scalar, vector and tensor potentials as the following [41]:

$$\Delta(r) = 4q_1 \frac{e^{-2\alpha r}}{(1-e^{-2\alpha r})^2} - q_2 \frac{(1+e^{-2\alpha r})}{(1-e^{-2\alpha r})} + \frac{v_0}{(1-e^{-2\alpha r})} - \frac{v_1}{(1-e^{-2\alpha r})^2} \tag{25}$$

$$U(r) = -\frac{H}{r}, \quad H = \frac{Z_f Z_g e^2}{4\pi\varepsilon_0}, \quad r \rangle R_c \tag{26}$$

Substituting Eq. (25) and Eq. (26) into Eq. (24), we obtain the lower radial equation of Dirac equation as:

$$\left\{ \frac{d^2}{dr^2} - \frac{\lambda_k(\lambda_k - 1)}{r^2} - \gamma' + \beta'\left( 4q_1 \frac{e^{-2\alpha r}}{(1-e^{-2\alpha r})^2} - q_2 \frac{(1+e^{-2\alpha r})}{(1-e^{-2\alpha r})} + \frac{v_0}{(1-e^{-2\alpha r})} - \frac{v_1}{(1-e^{-2\alpha r})^2} \right) \right\} G_{n_r,k}(r) = 0 \tag{27}$$

Where $\lambda_k = k + H$, $\gamma' = (M + E_{n,k})(M - E_{n,k} + C_{ps})$ and $\beta' = (M - E_{n,k} + C_{ps})$

By using the transformation $s = \exp(-2\alpha r)$ and employing the improved approximation Eq. (27) brings into the form:

$$G''_{n,k}(s) + \frac{(1-s)}{s(1-s)} G'_{n,k}(s) + \frac{1}{s^2(1-s)^2} \left[ A's^2 + B's + C' \right] G_{n,k}(s) = 0 \tag{28}$$

Where the parameters $A'$, $B'$ and $C'$ are considered as follows:

$$\begin{aligned} A' &= -\frac{1}{4\alpha^2}\left[\gamma' - \beta' q_2\right] \\ B' &= \frac{1}{4\alpha^2}\left[2\gamma' + 4\beta' q_1 - \beta' v_0\right] \\ C' &= -\frac{1}{4\alpha^2}\left[4\alpha^2 \lambda_k(\lambda_k - 1) + \gamma' + \beta' q_2 - \beta' v_0 + \beta' v_1\right] \end{aligned} \tag{29}$$

We can easily obtain the coefficients $k_i$ (i = 1, 2, 3) by comparing Eq. (28) with Eq. (1) as follows:

$k'_1 = k'_2 = k'_3 = 1$ \hfill (30)

The values of the coefficients $k'_i$ (i = 4, 5) are also found from Eq. (4) as below:

$$\begin{aligned} k'_4 &= \sqrt{-C'} \\ k'_5 &= \frac{1}{2} + \sqrt{\frac{1}{4} - [A' + B' + C']} \end{aligned} \tag{31}$$

We have Eq. (32) for energy Eigen-values by the use of Eq. (2):

$$\left[ \frac{-C' - \left[\frac{1}{2} + \sqrt{\frac{1}{4} - [A' + B' + C']}\right]^2 - \left[\frac{1-2n}{2} - \frac{1}{2}(1 - \sqrt{-4A'})\right]^2}{2\left[\frac{1-2n}{2} - \frac{1}{2}(1 - \sqrt{-4A'})\right]} \right]^2 - \left[\frac{1}{2} + \sqrt{\frac{1}{4} - [A' + B' + C']}\right]^2 = 0 \tag{32}$$

In Tables 4–6, we give the numerical results for the pseudo-spin symmetric energy Eigen-values (in units of fm$^{-1}$).

**Table 4.** The energy Eigen-values (in units of fm$^{-1}$) for the pseudo-spin symmetry limit in the presence and absence of Coulomb-like tensor interaction M=10 fm$^{-1}$, c=1, ℏ =1, α=0.05fm$^{-1}$, V$_1$=0.4fm$^{-1}$, V$_0$=0.3fm$^{-1}$, q$_1$=0.1fm$^{-1}$, q$_2$=-0.2fm$^{-1}$, C$_{ps}$ =-5fm$^{-1}$.

| l | n,k<0 | State (l,j) | E$^{ps}_{n,k}$ (H=0) | E$^{ps}_{n,k}$ (H=0.65) | n,k>0 | State (l+2,j+1) | E$^{ps}_{n,k}$ (H=0) | E$^{ps}_{n,k}$ (H=0.65) |
|---|---|---|---|---|---|---|---|---|
| 1 | 1,-1 | 1s$_{1/2}$ | -7.065764205 | -5.701645852 | 1,2 | 1d$_{3/2}$ | -7.065764205 | -7.916615587 |
| 2 | 1,-2 | 1p$_{3/2}$ | -8.242801196 | -7.570912051 | 1,3 | 1f$_{5/2}$ | -8.242801196 | -8.691210574 |
| 3 | 1,-3 | 1d$_{5/2}$ | -8.872276828 | -8.505344072 | 1,4 | 1g$_{7/2}$ | -8.872276828 | -9.132415328 |
| 4 | 1,-4 | 1f$_{7/2}$ | -9.241962415 | -9.022813407 | 1,5 | 1h$_{9/2}$ | -9.241962415 | -9.404940153 |
| 1 | 2;-1 | 2s$_{1/2}$ | -8.20892624 | -7.540097339 | 2,2 | 2d$_{3/2}$ | -8.20892624 | -8.656719162 |
| 2 | 2,-2 | 2p$_{3/2}$ | -8.837973611 | -8.470939351 | 2,3 | 2f$_{5/2}$ | -8.837973611 | -9.098960172 |
| 3 | 2,-3 | 2d$_{5/2}$ | -9.209115205 | -8.988907697 | 2,4 | 2g$_{7/2}$ | -9.209115205 | -9.37334647 |
| 4 | 2,-4 | 2f$_{7/2}$ | -9.44496483 | -9.303214658 | 2,5 | 2h$_{9/2}$ | -9.44496483 | -9.554705712 |

**Table 5.** The energy Eigen-values (in units of fm$^{-1}$) for the pseudo-spin symmetry limit with parameters M=10fm$^{-1}$, c=1, ℏ =1, V$_1$=4fm$^{-1}$, V$_0$=3fm$^{-1}$, q$_1$=1fm$^{-1}$, q$_2$=-2fm$^{-1}$, C$_{ps}$ =-5fm$^{-1}$.

| α(fm$^{-1}$) | E$^{ps}_{n,k}$(H=0) | | E$^{ps}_{n,k}$(H=0.65) | |
|---|---|---|---|---|
| | 1s$_{1/2}$ | 1d$_{3/2}$ | 1s$_{1/2}$ | 1d$_{3/2}$ |
| 0.3 | -4.313211747 | -4.313211747 | -2.304589779 | -5.851610879 |
| 0.35 | -5.441940815 | -5.441940815 | -3.496832686 | -6.846892358 |
| 0.4 | -6.330481794 | -6.330481794 | -4.501756572 | -7.592836004 |
| 0.5 | -7.58011028 | -7.58011028 | -6.029844453 | -8.587003959 |
| 0.6 | -8.366177606 | -8.366177606 | -7.07440217 | -9.176918108 |
| 0.7 | -8.871669319 | -8.871669319 | -7.792319679 | -9.537028876 |

**Table 6.** The energy Eigen-values (in units of fm$^{-1}$) for the pseudo-spin symmetry limit with parameters c=1, ℏ=1, α=0.4fm$^{-1}$, V$_1$=4fm$^{-1}$, V$_0$=3fm$^{-1}$, q$_1$=1fm$^{-1}$, q$_2$=-2fm$^{-1}$, C$_{ps}$ =-5fm$^{-1}$

| M(fm$^{-1}$) | E$^{ps}_{n,k}$(H=0) | | E$^{ps}_{n,k}$(H=0.65) | |
|---|---|---|---|---|
| | 1s$_{1/2}$ | 1d$_{3/2}$ | 1s$_{1/2}$ | 1d$_{3/2}$ |
| 5.5 | -4.282887209 | -4.282887209 | -3.435345049 | -4.891201159 |
| 6 | -4.519466496 | -4.519466496 | -3.562129097 | -5.200846041 |
| 6.5 | -4.751863076 | -4.751863076 | -3.685040453 | -5.506165891 |
| 7 | -4.981434942 | -4.981434942 | -3.80536706 | -5.808539775 |
| 8 | -5.435100161 | -5.435100161 | -4.04106051 | -6.407528709 |
| 9 | -5.884189891 | -5.884189891 | -4.27265172 | -7.001673282 |
| 10 | -6.330481794 | -6.330481794 | -4.501756572 | -7.592836004 |

By using Eq. (3) and Eq. (31) we can finally obtain the lower component of the Dirac spinor as below

$$G^{ps}_{n,k}(r) = N\left(e^{-2\alpha r}\right)^{\left(\sqrt{-c'}\right)} \left(1-e^{-2\alpha r}\right)^{\left(\frac{1}{2}+\sqrt{\frac{1}{4}+A'+B'+C'}\right)} {}_2F_1\left(-n, n+2\left(\sqrt{-c'}+\frac{1}{2}+\sqrt{\frac{1}{4}+A'+B'+C'}\right); 2\sqrt{-c'}+1, e^{-2\alpha r}\right)$$ (33)

Where N′ is the normalization constant, on the other hand, the upper component of the Dirac spinor can be calculated from Eq. (33) as:

$$F_{n,k}^{ps}(r) = \frac{1}{M - E_{n,k}^{ps} + C_{ps}} \left( \frac{d}{dr} - \frac{k}{r} + U(r) \right) G_{n,k}^{ps}(r) \quad (34)$$

We have obtained the energy Eigen-values and the spinors of the radial Dirac equation for Eckart plus Hulthen potentials with Coulomb-like tensor interaction in the presence of the pseudo-spin symmetry for k≠0. We show in Fig (2) behavior energy for various H in the spin and pseudo-spin symmetry.

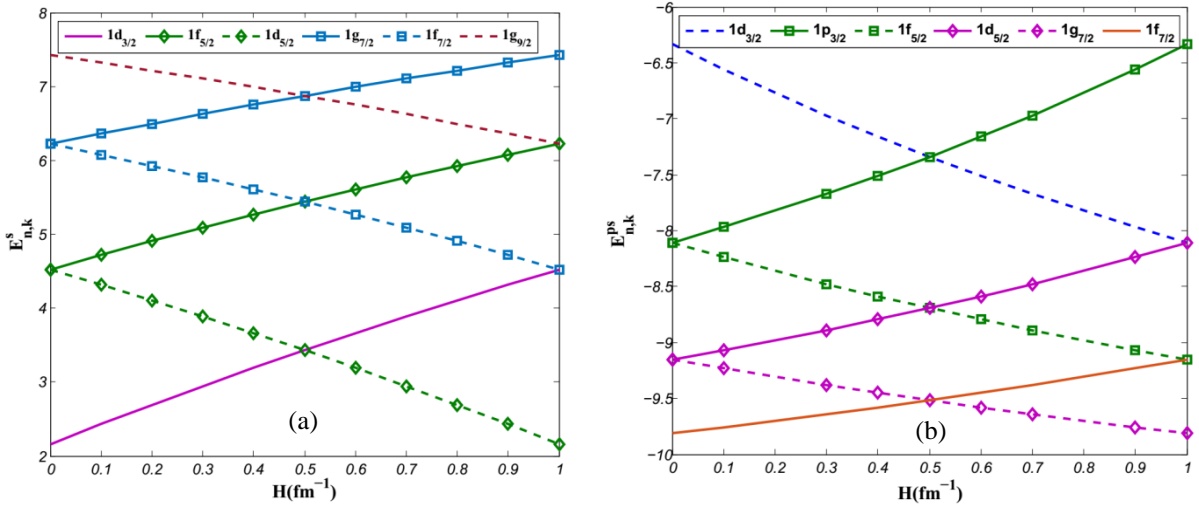

**Fig. 2.** Energy spectra in the (a)-spin and (b)-pseudo-spin symmetries at various H for Coulomb-tensor interaction with parameters M=10 fm⁻¹, c=1, ℏ =1, α=0.4fm⁻¹, V₁=4fm⁻¹, V₀=3fm⁻¹, q₁=1fm⁻¹, q₂ᵖˢ=-2fm⁻¹, q₂ˢ=2fm⁻¹, Cₛ =5fm⁻¹, Cₚₛ =-5fm⁻¹.

## 5.1. Solution Spin symmetry with Coulomb plus Yukawa-like tensor interaction

In this section for the spin symmetry, we consider $\Delta(r) = C_s$, $\Sigma(r)$ and $U(r)$ as the following:

$$\Sigma(r) = 4q_1 \frac{e^{-2\alpha r}}{(1-e^{-2\alpha r})^2} - q_2 \frac{(1+e^{-2\alpha r})}{(1-e^{-2\alpha r})} + \frac{v_0}{(1-e^{-2\alpha r})} - \frac{v_1}{(1-e^{-2\alpha r})^2} \quad (35)$$

$$U(r) = -\frac{H}{r} - A \frac{\exp(-2\alpha r)}{r} \quad (36)$$

Where H and A are the real parameters, substitution of Eq. (35) and (36) into Eq. (12) yields:

$$F_{n,k}''(s) + \frac{(1-s)}{s(1-s)} F_{n,k}'(s) + \frac{1}{s^2(1-s)^2} \left[ \chi_2 s^2 + \chi_1 s + \chi_0 \right] F_{n,k}(s) = 0 \quad (37)$$

Where the parameters χ₂, χ₁ and χ₀ are considered as the follows:

$$\chi_2 = -\frac{1}{4\alpha^2}[\gamma + \beta q_2] + A(A-1)$$

$$\chi_1 = \frac{1}{4\alpha^2}[2\gamma - 4\beta q_1 + \beta v_0] - 2A\tau_k \qquad (38)$$

$$\chi_0 = -\frac{1}{4\alpha^2}\left[4\alpha^2 \tau_k (\tau_k - 1) + \gamma - \beta q_2 + \beta v_0 - \beta v_1\right]$$

By comparing Eq. (37) with Eq. (1), we can easily obtain the coefficients $k_i$ (i = 1, 2, 3) as follows:

$k_1 = k_2 = k_3 = 1$ (39)

The values of the coefficients $k_i$ (i = 4, 5) are also found from Eq.(4) as below:

$$k_4 = \sqrt{-\chi_0}$$

$$k_5 = \frac{1}{2} + \sqrt{\frac{1}{4} - [\chi_2 + \chi_1 + \chi_0]} \qquad (40)$$

By the use of energy equation, Eq. (2) for energy Eigen-values we have:

$$\left[\frac{-\chi_0 - \left[\frac{1}{2} + \sqrt{\frac{1}{4} - [\chi_2 + \chi_1 + \chi_0]}\right]^2 - \left[\frac{1-2n}{2} - \frac{1}{2}\left(1 - \sqrt{-4\chi_2}\right)\right]^2}{2\left[\frac{1-2n}{2} - \frac{1}{2}\left(1 - \sqrt{-4\chi_2}\right)\right]}\right]^2 - \left[\frac{1}{2} + \sqrt{\frac{1}{4} - [\chi_2 + \chi_1 + \chi_0]}\right]^2 = 0 \quad (41)$$

In reference to Eq. (3) and using Eq. (40), we can obtain the upper wave function:

$$F_{n,k}^s(r) = N\left(e^{-2\alpha r}\right)^{\left(\sqrt{-\chi_0}\right)}\left(1 - e^{-2\alpha r}\right)^{\left(\frac{1}{2} + \sqrt{\frac{1}{4} + \chi_2 + \chi_1 + \chi_0}\right)} {}_2F_1\left(-n, n + 2\left(\sqrt{-\chi_0} + \frac{1}{2} + \sqrt{\frac{1}{4} + \chi_2 + \chi_1 + \chi_0}\right); 2\sqrt{-\chi_0} + 1, e^{-2\alpha r}\right) \quad (42)$$

Where N is the normalization constant, on the other hand, the upper component of the Dirac spinor can be calculated from Eq. (42) as:

$$G_{n,k}^s(r) = \frac{1}{M + E_{n,k}^s - C_s}\left(\frac{d}{dr} + \frac{k}{r} - U(r)\right)F_{n,k}^s(r) \qquad (43)$$

The effects of the Yukawa-like tensor interactions on the upper and lower components for the spin symmetry are shown in Figs. 3.

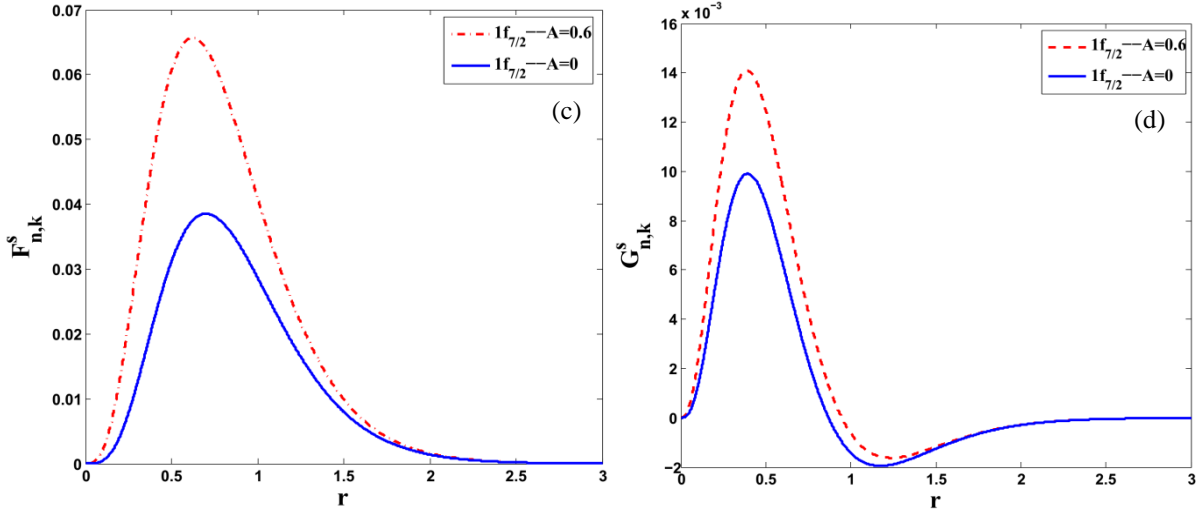

**Fig. 3.** (c)-upper and (d)-lower components of $1f_{7/2}$ in the spin symmetry in the presence and absence of Yukawa tensor interaction for M=10 fm$^{-1}$, c=1, ℏ =1, α=0.4fm$^{-1}$, $V_1$=4fm$^{-1}$, $V_0$=3fm$^{-1}$, $q_1$=1fm$^{-1}$, $q_2$=2fm$^{-1}$, $C_s$=5fm$^{-1}$.

We have obtained the energy Eigen-values and the spinors of the radial Dirac equation for Eckart plus Hulthen potentials with the spin symmetry for k≠0 in the presence and absence of Yukawa tensor.

### 5.2. Solution Pseudo-Spin symmetry with Coulomb plus Yukawa-like tensor interaction

In this section for the pseudo-spin symmetry, we consider $\sum(r) = C_s$, $\Delta(r)$ and U(r) as the following:

$$\Delta(r) = 4q_1 \frac{e^{-2\alpha r}}{(1-e^{-2\alpha r})^2} - q_2 \frac{(1+e^{-2\alpha r})}{(1-e^{-2\alpha r})} + \frac{v_0}{(1-e^{-2\alpha r})} - \frac{v_1}{(1-e^{-2\alpha r})^2} \tag{44}$$

$$U(r) = -\frac{H}{r} - A\frac{\exp(-2\alpha r)}{r} \tag{45}$$

Where H and A is the real parameter. Substitution of Eq. (44) and (45) into Eq. (24) yields:

$$G''_{n,k}(s) + \frac{(1-s)}{s(1-s)}G'_{n,k}(s) + \frac{1}{s^2(1-s)^2}\left[\chi'_2 s^2 + \chi'_1 s + \chi'_0\right]G_{n,k}(s) = 0 \tag{46}$$

Where the parameters $\chi'_2$, $\chi'_1$ and $\chi'_0$ are considered as the follows:

$$\chi'_2 = -\frac{1}{4\alpha^2}[\gamma - \beta q_2] + A(A+1)$$

$$\chi'_1 = \frac{1}{4\alpha^2}[2\gamma + 4\beta q_1 - \beta v_0] - 2A(\lambda_k - 1) \tag{47}$$

$$\chi'_0 = -\frac{1}{4\alpha^2}\left[4\alpha^2 \lambda_k (\lambda_k - 1) + \gamma + \beta q_2 - \beta v_0 + \beta v_1\right]$$

By comparing Eq. (57) with Eq. (2), we can easily obtain the coefficients k′$_i$ (i = 1, 2, 3) as follows:

k′$_1$=k′$_2$= k′$_3$=1 (48)

The values of the coefficients $k'_i$ (i = 4, 5) are also found from Eq. (4) as below:

$$k'_4 = \sqrt{-\chi'_0}$$
$$k'_5 = \frac{1}{2} + \sqrt{\frac{1}{4} - [\chi'_2 + \chi'_1 + \chi'_0]} \tag{49}$$

By using the energy equation, Eq. (2) for energy Eigen-values we have:

$$\left[\frac{-\chi'_0 - \left[\frac{1}{2} + \sqrt{\frac{1}{4} - [\chi'_2 + \chi'_1 + \chi'_0]}\right]^2 - \left[\frac{1-2n}{2} - \frac{1}{2}(1 - \sqrt{-4\chi'_2})\right]^2}{2\left[\frac{1-2n}{2} - \frac{1}{2}(1 - \sqrt{-4\chi'_2})\right]} - \left[\frac{1}{2} + \sqrt{\frac{1}{4} - [\chi'_2 + \chi'_1 + \chi'_0]}\right]\right]^2 = 0 \tag{50}$$

By the use of Eq. (3) and Eq. (49) we can finally obtain the lower component of the Dirac spinor as below

$$G_{n,k}^{ps}(r) = N\left(e^{-2\alpha r}\right)^{\left(\sqrt{-\chi'_0}\right)}\left(1-e^{-2\alpha r}\right)^{\left(\frac{1}{2}+\sqrt{\frac{1}{4}+\chi'_2+\chi'_1+\chi'_0}\right)}{}_2F_1\left(-n, n+2\left(\sqrt{-\chi'_0} + \frac{1}{2} + \sqrt{\frac{1}{4}+\chi'_2+\chi'_1+\chi'_0}\right); 2\sqrt{-\chi'_0}+1, e^{-2\alpha r}\right) \tag{51}$$

Where N′ is the normalization constant, on the other hand, the upper component of the Dirac spinor can be calculated from Eq. (51) as:

$$F_{n,k}^{ps}(r) = \frac{1}{M - E_{n,k}^{ps} + C_{ps}}\left(\frac{d}{dr} - \frac{k}{r} + U(r)\right)G_{n,k}^{ps}(r) \tag{52}$$

The effects of the Yukawa-like tensor interactions on the upper and lower components for the pseudo-spin symmetry are shown in Figs. 4.

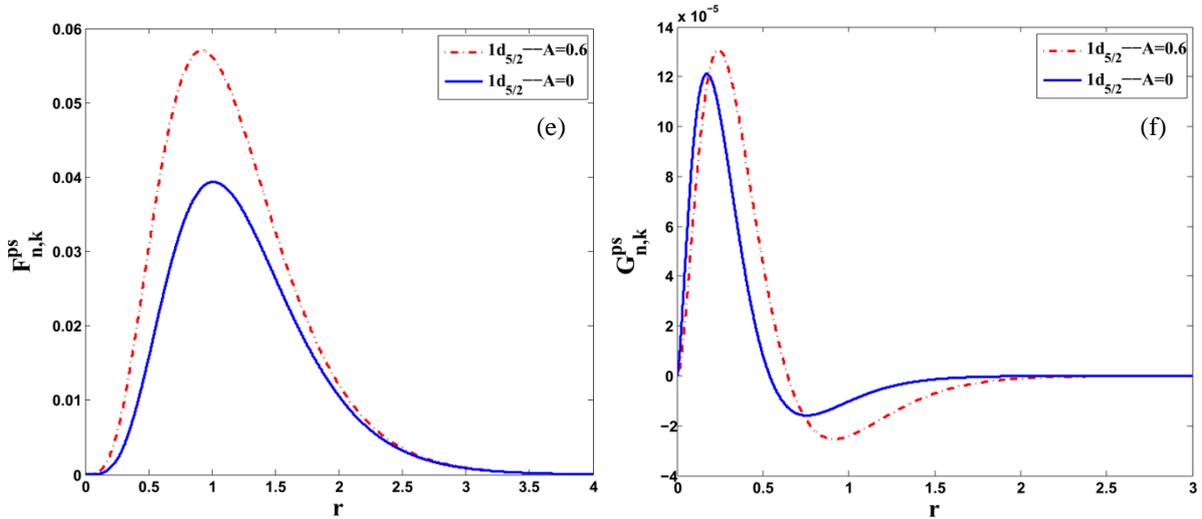

**Fig. 4.** (e)-upper and (f)-lower components of $1d_{5/2}$ in the pseudo-spin symmetry in the presence and absence of Yukawa tensor interaction for M=10 fm$^{-1}$, c=1, ℏ=1, α=0.4fm$^{-1}$, $V_1$=4fm$^{-1}$, $V_0$=3fm$^{-1}$, $q_1$=1fm$^{-1}$, $q_2$=-2fm$^{-1}$, $C_{ps}$=-5fm$^{-1}$.

We have obtained the energy Eigen-values and the spinors of the radial Dirac equation for Eckart plus Hulthen potentials with the pseudo-spin symmetry for k≠0 in the presence and absence of Yukawa tensor interaction. In Figs. 5 and 6 we show the effect of Coulomb-like tensor interaction and Yukawa plus Coulomb-like tensor interaction in the remove of degeneracy.

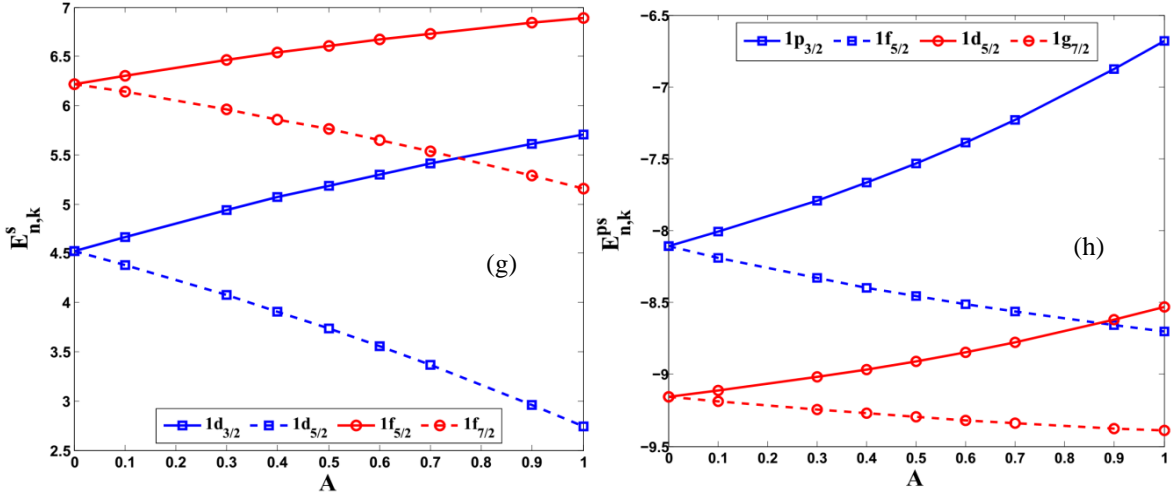

**Fig. 5.** Energy spectra in the (g)-spin and (h)-pseudo-spin symmetries versus A for Yukawa-like tensor interaction with parameters M=10 fm$^{-1}$, c=1, $\hbar$ =1, α=0.4fm$^{-1}$, $V_1$=4fm$^{-1}$, $V_0$=3fm$^{-1}$, $q_1$=1fm$^{-1}$, $q_2^{ps}$=-2fm$^{-1}$, $q_2^s$=2fm$^{-1}$, $C_s$ =5fm$^{-1}$, $C_{ps}$ =-5fm$^{-1}$.

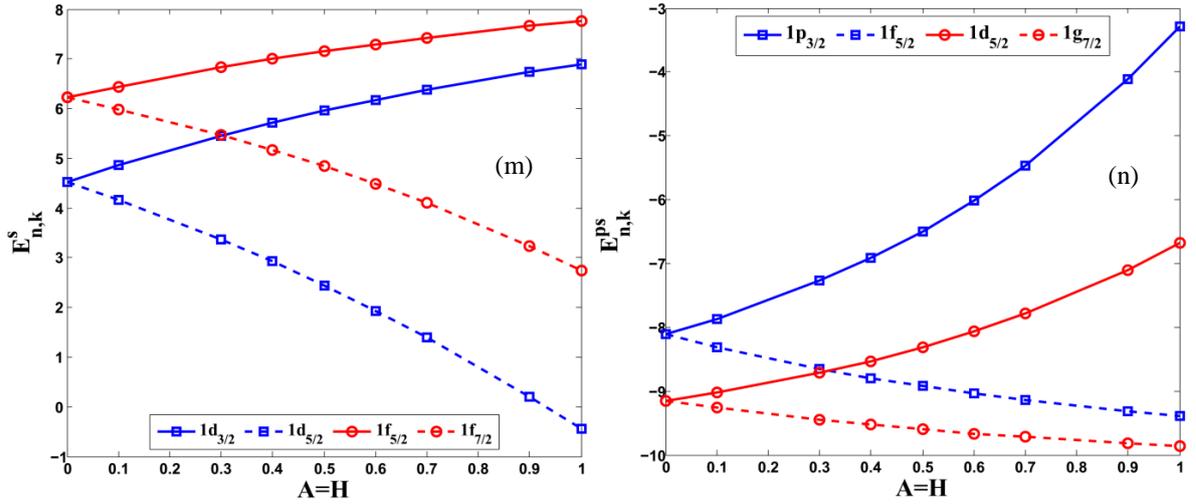

**Fig. 6.** Energy spectra in the (m) spin and (n) pseudo-spin symmetries versus A=H for Yukawa plus Coulomb like tensor interaction with parameters M=10 fm$^{-1}$, c=1, $\hbar$ =1, α=0.4fm$^{-1}$, $V_1$=4fm$^{-1}$, $V_0$=3fm$^{-1}$, $q_1$=1fm$^{-1}$, $q_2^{ps}$=-2fm$^{-1}$, $q_2^s$=2fm$^{-1}$, $C_s$ =5fm$^{-1}$, $C_{ps}$ =-5fm$^{-1}$.

In Fig. 7 we show the comparing between Coulomb and Yukawa like tensor interaction for spin and pseudo-spin symmetries.

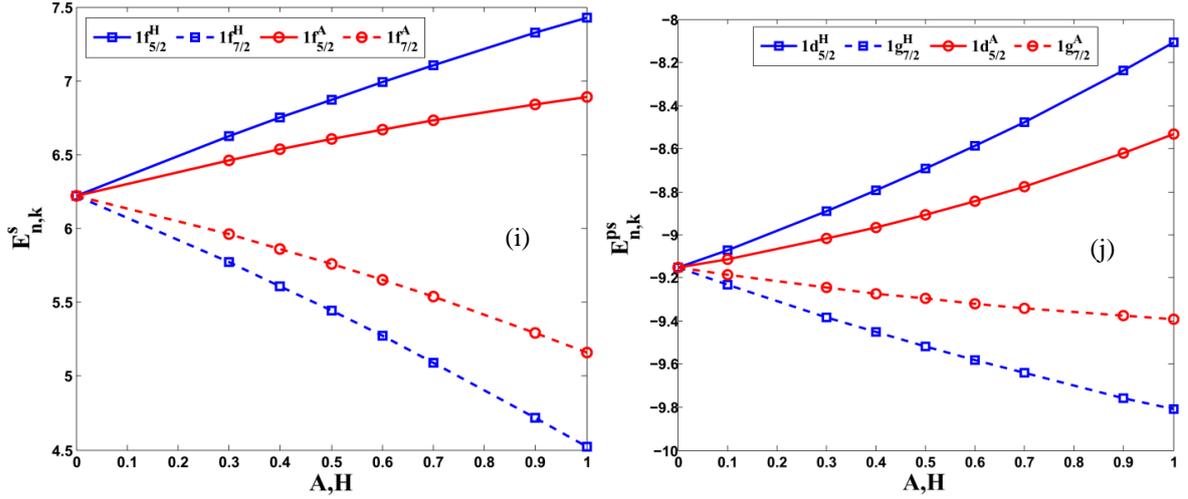

**Fig. 7.** Energy spectra in the (i)-spin and (j)-pseudo-spin symmetries at various H and A for compare between Coulomb and Yukawa-like tensor interaction with parameters M=10 fm$^{-1}$, c=1, ℏ =1, α=0.4fm$^{-1}$, V$_1$=4fm$^{-1}$, V$_0$=3fm$^{-1}$, q$_1$=1fm$^{-1}$, q$_2^{ps}$=-2fm$^{-1}$, q$_2^s$=2fm$^{-1}$, C$_s$ =5fm$^{-1}$, C$_{ps}$ =-5fm$^{-1}$.

In Fig.5, 6 and 7 we showed that degeneracy is removed by tensor interaction. Furthermore, the amount of the energy difference between the two states in the doublets increases with increasing H and A.

## 6. Results and Discussion

We obtained the energy Eigen-values in the absence and the presence of the Coulomb-like tensor potential for various values of the quantum numbers n and k. In table 1 and 4 in the absence of the tensor interaction (H = 0), the degeneracy between spin doublets and pseudo-spin doublets are observed. For example, we observe the degeneracy in (1p$_{1/2}$, 1p$_{3/2}$), (1d$_{3/2}$, 1d$_{5/2}$)..., etc in the spin symmetry, and we observe the degeneracy in (1s$_{1/2}$, 1d$_{3/2}$), (1p$_{3/2}$, 1f$_{5/2}$)..., etc in the pseudo-spin symmetry. When we consider the tensor interaction for example by parameter H=0.65, the degeneracy is removed. In table 2 and 3 for the spin symmetry also in table 5 and 6 for the pseudo-spin symmetry, we show that degeneracy exist between spin doublets for several of parameters α and M, and we show that degeneracy is removed in the present of tensor interaction. In Fig (2) the degeneracy is removed by tensor interaction effect in spin symmetry and pseudo-spin symmetry also the amount of the energy difference between the two states in the doublets increases with increasing parameter H. The effects of the Yukawa-like tensor interactions on the upper and lower components of radial Dirac equation for the symmetries are shown in Figs. 3 and 4. The sensitiveness of the pseudo-spin doublets (1p$_{3/2}$, 1f$_{5/2}$) and (1d$_{5/2}$, 1g$_{7/2}$) and spin doublets (1f$_{5/2}$, 1f$_{7/2}$) and (1d$_{3/2}$, 1d$_{5/2}$) for the effects of the Yukawa-like tensor interactions and for Yukawa plus Coulomb-like tensor interaction are given in Figs. 5 and 6. In Fig. 7 we show that the coulomb tensor interaction is stronger than Yukawa-like tensor interaction for remove degeneracy.

## 7. Conclusions

In this paper, we have discussed approximately the solutions of the Dirac equation for Eckart plus Hulthen potentials with Spin Symmetry and Pseudo-spin Symmetry for $k \neq 0$. We obtained the energy Eigen-values and the wave function in terms of the generalized polynomials functions via the Formula method. To show the accuracy of the present model, some numerical values of the energy levels are shown in figure 3, 4, 5 and 6. We have showed that the energy degeneracy in pseudo-spin and spin doublets is removed by the tensor interaction effect.

## Acknowledgements

We would like to thank the kind referee for his helpful comments and suggestions which have improved the manuscript greatly.## References